\documentclass[11pt]{article}
\usepackage[utf8]{inputenc}
\usepackage{geometry}
\renewcommand\[{\begin{equation}} \renewcommand\]{\end{equation}}
\usepackage{setspace}
\usepackage[font={onehalfspacing, it}]{caption}
\usepackage[margin=10pt,font={footnotesize},labelfont=bf,labelsep=colon]{caption}
\usepackage{geometry} % you need this for yalephd.cls to work.
\usepackage{graphicx} % you probably want the rest of these.
\usepackage{dcolumn}
\usepackage{bm}
\usepackage{amsmath}
\usepackage{amsfonts}
\usepackage{amssymb}
\usepackage{appendix}
\usepackage{comment}
\usepackage{hyperref}
\hypersetup{colorlinks=true,linkcolor=black,citecolor=blue, urlcolor=cyan}

\begin{document}
\title{Case for the double-blind peer review}

\author{Lucie Tvrznikova}

\maketitle
Peer review is a process designed to produce a fair assessment of
research quality prior to the publication of scholarly work in a journal.
The process consists of sending a candidate work such as a journal
article to a number of similarly qualified professionals for feedback
and informs publishers of academic journals whether a work should
be accepted or rejected. Peer review is used as a gold standard in
assessing manuscripts and is essential for academics since tenure,
grant applications, and many other career-related decisions are based
upon the quality of their work and publications making the stakes
are high. Most journals today use a single-blind review method, in
which the reviewer's identity is concealed to the author of the article,
but not vice versa. However, some journals use alternative review
strategies~\cite{blank1991effects,baggs2008blinding}, such as double-blind
review, where both the author's and the reviewer's identities are
concealed, or open review~\cite{Hopewell_2014,nature_openReview},
where all information is public. 

It is well known that implicit biases influence people's decision
making but ideally decisions regarding the importance of scientific
contributions would be free of these biases. Nevertheless, demographics,
nepotism, and seniority have been shown to affect reviewer behavior
suggesting the most common, single-blind review method (or the less
common open review method) might be biased~\cite{nature_nepoticm,Link_1998}.
Blinding both the author's and reviewer's identity can reduce exposure
to extraneous information thus bypassing bias and optimizing decision
making. Although this article focuses on academic journals, blinding
has the potential to affect decision making in many areas: it has
been demonstrated that blinding applicants in the first stages of
grant applications reduces biases related to researcher\textquoteright s
characteristics~\cite{solans2017blinding} and blind musical auditions
have significantly increased the number of women in symphony orchestras~\cite{goldin2000orchestrating}.
Moss \textit{et al.} also showed that blinding the gender of job applicants
can reduce unconscious and unintended bias~\cite{moss2012science}. 

Many characteristics can be subject to biases that might skew the
reviewer\textquoteright s decisions as discussed below. For example,
there is substantial evidence for biases in favor of famous authors
and world-renown institutions, but there are only limited data on
the impact of review on women and minorities. Budden \textit{et al.}
found that switching to a double-blind review in an ecology journal
led to a small increase of female first author paper~\cite{budden2008double}.
However, these findings were later rebutted~\cite{whittaker2008journal,webb2008does,engqvist2008double}
and this increase was attributed to a proportional increase of women
in the workforce overall. Journal of Neurophysiology also found no
gender bias in their publications, but carefully notes that at the
time of the analysis 5 out of 9 members of the editorial panel were
women and only 191 of 713 submissions (26\%) were from women-first
authors~\cite{lane2009there}. Blank found that although women perform
slightly better under a double-blind system, the data were statistically
not significant~\cite{blank1991effects}. It should be noted that
women published only 150 out of 1,223 analyzed papers (12\%). Other
journals also found author identity did not influence acceptance rates
or quality ratings in review~\cite{chung2015double,borsuk2009name}.

Conversely, after the Modern Language Association (MLA) switched to
double-blind review a substantial increase in acceptances for female-authored
publications was observed, eventually comparable to that for men~\cite{largent2016blind}.
Tomkins \textit{et al.} did not find any gender bias in their study
of the review process in computer science using their data alone,
but combining their results with previous studies showed a statistically
significant gender effect bias~\cite{tomkins2017single}. Studying
this topic, Helmer \textit{et al.} demonstrated that women are currently
underrepresented in the peer review process and that editors of both
genders preferred same-gendered authors~\cite{helmer2017gender}.
 This behavior highlights the need to employ review methods that combat
subtler forms of gender bias in scholarly publishing.

Despite the mixed data regarding the effect of double-blind review
on gender bias, it has been demonstrated that double-blind review
can mitigate biases arising from researcher's popularity or location.
Tomkins \textit{et al.} found that single-blind reviewers were significantly
more likely to recommend for acceptance papers from famous authors
and top institutions~\cite{tomkins2017single}. Moreover, if reviewers
knew the author's identity, they disfavored authors that were not
sufficiently embedded in their research community~\cite{seeber2017does}.
Link discovered that in a medical journal the location of the authors
mattered as US reviewers ranked US papers much more favorably compared
to the non-US ones~\cite{Link_1998}. Additionally, Blank found that
under a double-blind review acceptance rates were lower and reviewers
were more critical~\cite{blank1991effects}. This suggests that a
double-blind system might lead to more critical feedback from reviewers
and a mix of accepted papers from more diverse authorship. 

It should be noted that before a manuscript reaches the review stage,
journal editors who have access to authors' information reject many,
sometimes most, of the papers submitted to a journal before a reviewer
can see them. While motives will differ between editors and reviewers,
there is no reason to assume that editors are less susceptible to
bias than reviewers~\cite{garvalov2015stands}. This issue could
be overcome with the use of double-blind review since editors would
also lose access to authorship information.

A common concern for double-blind review is the identification of
author or institution from self-citations, nature of work, or personal
connections~\cite{Yankauer_1991,katz2002incidence}. Hill and Provost
found that even using the best method to identify authors based on
discriminative self-citations, authors were identified correctly only
40-45\% of the time suggesting that even in the worst-case scenario
blinding is successful most of the time~\cite{Hill_2003}. Other
arguments against double-blind review such as administrative inconvenience,
posting articles to other websites prior to their publications, possible
conflicts of interest, or tradition can be overcome~\cite{goues2017effectiveness}.
In physics, the issue of identification becomes significantly more
relevant due to the high number of large international collaborations
of various scales ranging from dozens to thousands of members. Often,
publications from these collaborations will be trivial to identify.
However, it is also unlikely they will suffer any disadvantages while
the double-blind review could still benefit researchers who might
currently be affected by the biases in the review process.

A scientist's career should not be influenced by stereotypes, but
rather depend on an individual's proven track record to perform. Reviewers
are people too, and since biases can affect anyone the community should
strive toward a peer review system that makes its best effort to overcome
the susceptibility to bias. Current research suggests that double-blind
review reduces bias that might arise while assessing the quality of
the research presented in a manuscript, without imposing any major
downsides. Double-blind review offers a solution to many biases stemming
from author's gender, seniority, and institution and should be a foundation
of any journal that strives to evaluate author's work strictly based
on the quality of the research presented in the manuscript.

\bibliographystyle{JHEP}
\bibliography{doubleBlind_review}

\providecommand{\href}[2]{#2}\begingroup\raggedright\begin{thebibliography}{10}

\bibitem{blank1991effects}
R.~M. Blank, \emph{{The Effects of Double-Blind versus Single-Blind Reviewing:
  Experimental Evidence from The American Economic Review}}, {\emph{American
  Economic Review} {\bfseries 81} (December, 1991) 1041--1067},
  [\href{https://arxiv.org/abs/https://ideas.repec.org/a/aea/aecrev/v81y1991i5p1041-67.html}{{\ttfamily
  https://ideas.repec.org/a/aea/aecrev/v81y1991i5p1041-67.html}}].

\bibitem{baggs2008blinding}
J.~G. Baggs, M.~E. Broome, M.~C. Dougherty, M.~C. Freda and M.~H. Kearney,
  \emph{Blinding in peer review: the preferences of reviewers for nursing
  journals},
  \href{http://dx.doi.org/10.1111/j.1365-2648.2008.04816.x}{\emph{Journal of
  Advanced Nursing} {\bfseries 64} 131--138}.

\bibitem{Hopewell_2014}
S.~Hopewell, G.~S. Collins, I.~Boutron, L.-M. Yu, J.~Cook, M.~Shanyinde et~al.,
  \emph{Impact of peer review on reports of randomised trials published in open
  peer review journals: retrospective before and after study},
  \href{http://dx.doi.org/10.1136/bmj.g4145}{\emph{BMJ} {\bfseries 349} (Jul,
  2014) g4145--g4145}.

\bibitem{nature_openReview}
\emph{Overview: Nature's peer review trial},
  \href{https://arxiv.org/abs/https://doi.org/10.1038/nature05535}{{\ttfamily
  https://doi.org/10.1038/nature05535}}.

\bibitem{nature_nepoticm}
C.~Wenner{\aa}s and A.~Wold, \emph{Nepotism and sexism in peer-review},
  \href{http://dx.doi.org/10.1038/387341a0}{\emph{Nature} {\bfseries 387} (May,
  1997) 341--343}.

\bibitem{Link_1998}
A.~M. Link, \emph{{US and Non-US Submissions}},
  \href{http://dx.doi.org/10.1001/jama.280.3.246}{\emph{JAMA} {\bfseries 280}
  (Jul, 1998) 246}.

\bibitem{solans2017blinding}
M.~Solans-Dom{\`e}nech, I.~Guillam{\'o}n, A.~Ribera, I.~Ferreira-Gonz{\'a}lez,
  C.~Carrion, G.~Permanyer-Miralda et~al., \emph{Blinding applicants in a
  first-stage peer-review process of biomedical research grants: An
  observational study},
  \href{http://dx.doi.org/10.1093/reseval/rvx021}{\emph{Research Evaluation}
  {\bfseries 26} (2017) 181--189}.

\bibitem{goldin2000orchestrating}
C.~Goldin and C.~Rouse, \emph{Orchestrating impartiality: The impact of "blind"
  auditions on female musicians},
  \href{http://dx.doi.org/10.3386/w5903}{\emph{American Economic Review}
  {\bfseries 90} (2000) 715--741}.

\bibitem{moss2012science}
C.~A. Moss-Racusin, J.~F. Dovidio, V.~L. Brescoll, M.~J. Graham and
  J.~Handelsman, \emph{Science faculty{\textquoteright}s subtle gender biases
  favor male students},
  \href{http://dx.doi.org/10.1073/pnas.1211286109}{\emph{Proceedings of the
  National Academy of Sciences} {\bfseries 109} (2012) 16474--16479}.

\bibitem{budden2008double}
A.~E. Budden, T.~Tregenza, L.~W. Aarssen, J.~Koricheva, R.~Leimu and C.~J.
  Lortie, \emph{Double-blind review favours increased representation of female
  authors},
  \href{http://dx.doi.org/https://doi.org/10.1016/j.tree.2007.07.008}{\emph{Trends
  in Ecology \& Evolution} {\bfseries 23} (2008) 4 -- 6}.

\bibitem{whittaker2008journal}
R.~J. Whittaker, \emph{Journal review and gender equality: a critical comment
  on budden et al.},
  \href{http://dx.doi.org/https://doi.org/10.1016/j.tree.2008.06.003}{\emph{Trends
  in Ecology \& Evolution} {\bfseries 23} (2008) 478 -- 479}.

\bibitem{webb2008does}
T.~J. Webb, B.~O'Hara and R.~P. Freckleton, \emph{Does double-blind review
  benefit female authors?},
  \href{http://dx.doi.org/https://doi.org/10.1016/j.tree.2008.03.003}{\emph{Trends
  in Ecology \& Evolution} {\bfseries 23} (2008) 351 -- 353}.

\bibitem{engqvist2008double}
L.~Engqvist and J.~G. Frommen, \emph{Double-blind peer review and gender
  publication bias},
  \href{http://dx.doi.org/https://doi.org/10.1016/j.anbehav.2008.05.023}{\emph{Animal
  Behaviour} {\bfseries 76} (2008) e1 -- e2}.

\bibitem{lane2009there}
J.~A. Lane and D.~J. Linden, \emph{{Is There Gender Bias in the Peer Review
  Process at Journal of Neurophysiology?}},
  \href{http://dx.doi.org/10.1152/jn.00196.2009}{\emph{Journal of
  Neurophysiology} {\bfseries 101} (2009) 2195--2196}.

\bibitem{chung2015double}
K.~C. Chung, M.~J. Shauver, S.~Malay, L.~Zhong, A.~Weinstein and R.~J. Rohrich,
  \emph{{Is double-blinded peer review necessary? The effect of blinding on
  review quality}},
  \href{http://dx.doi.org/10.1097/PRS.0000000000001820}{\emph{Plastic and
  reconstructive surgery} {\bfseries 136} (2015) 1369--1377}.

\bibitem{borsuk2009name}
R.~M. Borsuk et~al., \emph{To name or not to name: The effect of changing
  author gender on peer review},
  \href{http://dx.doi.org/10.1525/bio.2009.59.11.10}{\emph{BioScience}
  {\bfseries 59} (2009) 985--989}.

\bibitem{largent2016blind}
E.~A. Largent and R.~T. Snodgrass, \emph{Blind peer review by academic
  journals}, {\emph{Blinding as a Solution to Bias: Strengthening Biomedical
  Science, Forensic Science, and Law} (2016) 75--95}.

\bibitem{tomkins2017single}
A.~Tomkins, M.~Zhang and W.~D. Heavlin, \emph{Single versus double blind
  reviewing at wsdm 2017},  \href{https://arxiv.org/abs/1702.00502}{{\ttfamily
  1702.00502}}.

\bibitem{helmer2017gender}
M.~Helmer, M.~Schottdorf, A.~Neef and D.~Battaglia, \emph{Gender bias in
  scholarly peer review},
  \href{http://dx.doi.org/10.7554/eLife.21718}{\emph{Elife} {\bfseries 6}
  (2017) }.

\bibitem{seeber2017does}
M.~Seeber and A.~Bacchelli, \emph{Does single blind peer review hinder
  newcomers?},
  \href{http://dx.doi.org/10.1007/s11192-017-2264-7}{\emph{Scientometrics}
  {\bfseries 113} (Oct, 2017) 567--585}.

\bibitem{garvalov2015stands}
B.~K. Garvalov, \emph{Who stands to win from double-blind peer review?},
  \href{http://dx.doi.org/10.3402/arb.v2.26879}{\emph{Advances in Regenerative
  Biology} {\bfseries 2} (2015) 26879}.

\bibitem{Yankauer_1991}
A.~Yankauer, \emph{How blind is blind review?},
  \href{http://dx.doi.org/10.2105/ajph.81.7.843}{\emph{American Journal of
  Public Health} {\bfseries 81} (Jul, 1991) 843--845}.

\bibitem{katz2002incidence}
D.~S. Katz, A.~V. Proto and W.~W. Olmsted, \emph{Incidence and nature of
  unblinding by authors: our experience at two radiology journals with
  double-blinded peer review policies},
  \href{http://dx.doi.org/10.2214/ajr.179.6.1791415}{\emph{American Journal of
  Roentgenology} {\bfseries 179} (2002) 1415--1417}.

\bibitem{Hill_2003}
S.~Hill and F.~Provost, \emph{The myth of the double-blind review?},
  \href{http://dx.doi.org/10.1145/980972.981001}{\emph{ACM SIGKDD Explorations
  Newsletter} {\bfseries 5} (Dec, 2003) 179}.

\bibitem{goues2017effectiveness}
C.~L. Goues, Y.~Brun, S.~Apel, E.~Berger, S.~Khurshid and Y.~Smaragdakis,
  \emph{Effectiveness of anonymization in double-blind review},
  \href{http://dx.doi.org/10.1145/3208157}{\emph{Commun. ACM} {\bfseries 61}
  (May, 2018) 30--33}.

\end{thebibliography}\endgroup

\end{document}